\documentclass[reprint,aps,prl,showpacs,showkeys]{revtex4-1}

\usepackage{natbib}
\usepackage{graphics}
\usepackage{bm}
\usepackage{hyperref}

\begin{document}

\title{Exact self-similar solutions in Born--Infeld theory}

\author{E.~Yu.~Petrov}
\author{A.~V.~Kudrin}
\email{kud@rf.unn.ru}
\affiliation{Department of Radiophysics, University of Nizhny Novgorod,
23 Gagarin Ave., Nizhny Novgorod 603950, Russia}
\begin{abstract}
We present a new class of exact self-similar solutions possessing cylindrical
or spherical symmetry in Born--Infeld theory. A cylindrically symmetric
solution describes the propagation of a cylindrical electromagnetic
disturbance in a constant background magnetic field in Born--Infeld
electrodynamics. We show that this solution corresponds to vacuum
breakdown and the subsequent propagation of an electron--positron avalanche.
The proposed method of finding exact analytical solutions can be
generalized to the model of a spherically symmetric scalar Born--Infeld
field in the ($n+1$)-dimensional Minkowski space-time. As an example,
the case $n=3$ is discussed.
\end{abstract}

\pacs{03.50.Kk, 03.65.Pm}

\maketitle

Born--Infeld electrodynamics was proposed in the 1930s as a nonlinear generalization
of Maxwell electromagnetism~\cite{Bor}. The central idea of Born and Infeld
was to create a relativistic field theory which could admit a finite energy classical
solution describing an elementary electric charge. Since the appearance of quantum
electrodynamics (QED), the interest in classical field theory has faded considerably.
However, in that time, Born--Infeld (BI) theory received a fairly
unexpected development. It was related to Heisenberg's suggestion to describe multiple
$\pi$-meson emission in high-energy hadronic collisions as an expansion
of a nonlinear wave packet in the context of the effective scalar BI field~\cite{Hei}.
This classical Heisenberg's model has proven to be very fruitful and continues
to be developed in the physics of quark--gluon plasma~\cite{Kan,Pav}.

In the past decades, since its rediscovery in low energy limit of string
theories~\cite{Fra,Met,Cal,Gib1,Gib2,Gib3}, BI electrodynamics has been studied
extensively by a large number of workers (see, e.g.,~\cite{Gib1,Gib2,Gib3,Gal,Kie1,Kie2,Car,Chr,Ker} and references therein).
A substantial degree of interest in the subject has also been stimulated by
recent advances in laser technology. Modern high-power lasers make it possible to
reach an intensity level of $10^{26}$--$10^{28}$ W/cm$^2$. The corresponding
field strength is sufficient to make nonlinear electrodynamic effects in
vacuum measurable~\cite{Mou,Fed1,Tom}. The experimental verification of
BI electrodynamics could serve as an important evidence in favor of string
theory, and various ways for laboratory testing of the corresponding
nonlinear effects are currently discussed in the
literature~\cite{Tom,Den1,Den2,Fer1,Mun}.

It is known that BI electrodynamics has unique properties (causal propagation
and the absence of birefringence~\cite{Boi1,Des,Bia}) among other relativistic nonlinear
theories of the electromagnetic field. Born--Infeld theory possesses an impressive mathematical
beauty and has already found numerous applications in different branches of physics.

Because of a rather complicated nonlinearity of the BI field equations,
only several exact solutions are known in this theory. These are the
point charge solution found by the creators of the theory~\cite{Bor},
2D electrostatic solutions~\cite{Pry,Fer2}, and plane wave
solutions~\cite{Schr,Bar,Boi2,Aie}. The problem of finding new exact solutions
in BI electrodynamics is very topical. It is well known that BI equations
admit the existence of exact static singular solutions with finite total energy
(the so-called BIon solutions). The simplest cases are cylindrically or
spherically symmetric solutions with a singularity on the axis or at the
origin, respectively. Recently, such solutions has received much attention in
string or M theory~\cite{Cal,Gib2}. It should be noted, however, that the
properties of nonstationary solutions, which describe the propagation of
cylindrical and spherical wave disturbances, remain poorly studied not only in BI theory,
but also in Maxwell electrodynamics of nonlinear media. Some exact axisymmetric
solutions of the Maxwell equations in a nonlinear medium have recently been
found in~\cite{Pet1,Esk,Pet2}. In this work, we obtain exact self-similar
solutions possessing cylindrical or spherical symmetry in BI theory.

The Lagrangian density of BI electrodynamics is given by
\begin{equation}
L=(4\pi)^{-1} b^2\left(1-\sqrt{1-b^{-2}I-b^{-4}J^{2}}\right),\label{eq1}
\end{equation}
where $b$ is Born's constant~\cite{Bor},
$I=-\frac{1}{2}F_{\mu \nu}F^{\mu \nu}={\bf E}^{2}-{\bf B}^{2}$ and
$J=-\frac{1}{4}F_{\mu\nu} {\cal F}^{\mu\nu}={\bf E}\cdot{\bf B}$ are the Poincar\'{e}
invariants, and $F_{\mu\nu}$ and ${\cal F}^{\mu\nu}$ are the electromagnetic-field tensor
and the dual tensor, respectively. The electromagnetic field equations following
from Lagrangian density~(\ref{eq1}) formally coincide with the Maxwell equations if
the constitutive relations have the form ${\bf D}=4\pi\,{\partial L}/{\partial {\bf E}}$ and
${\bf H}=-4\pi\,{\partial L}/{\partial {\bf B}}$.
We introduce a cylindrical coordinate system ($r$, $\phi$, $z$) and assume that
the fields are independent of
$\phi$ and $z$. Then the BI equations admit solutions in which
only the $E_z$ and $B_\phi$ components are nonzero.
Denoting these components as $E$ and $B$, respectively, and using the fact
that $J$ vanishes in this case,
we can write the field equations in the form
\begin{equation}
\partial_{r}H+r^{-1}H=\partial_{\tau}D,\quad
\partial_{r}E=\partial_{\tau}B,\label{eq2}
\end{equation}
where $D=E(1-b^{-2}I)^{-1/2}$, $H=B(1-b^{-2}I)^{-1/2}$, $I=E^{2}-B^{2}$,
$\tau=ct$, and $c$ is the speed of light. The second equation in system~(\ref{eq2}) is
satisfied by putting
\begin{equation}
E=-b\,\partial_{\tau}\psi,\quad B=-b\,\partial_{r}\psi,\label{eq3}
\end{equation}
where $\psi$ is the normalized (to $b$) $z$-component of the vector potential. Assuming
that $I\neq b^{2}$, from the first equation in system~(\ref{eq2}) we obtain
\begin{eqnarray}
&&
[1-(\partial_{\tau}\psi)^{2}]\partial^{2}_{r}\psi+2\partial_{r}\psi\,
\partial_{\tau}\psi\,\partial_{r}(\partial_{\tau}\psi)
-[1+(\partial_{r}\psi)^{2}]\partial^{2}_{\tau}\psi\nonumber\\
&&
+(n-1)r^{-1}
[1-(\partial_{\tau}\psi)^{2}+(\partial_{r}\psi)^{2}]\partial_{r}\psi=0,\label{eq4}
\end{eqnarray}
where $n=2$ in the case considered. Here, the integer parameter $n$ is introduced for the following
reasons. Equation~(\ref{eq4}) can be considered as the Euler--Lagrange equation
for the scalar field $\psi(r,t)$ in the ($n+1$)-dimensional Minkowski space-time,
which follows from the action
\begin{equation}
S=\int [1-(\partial_{\tau}\psi)^{2}+(\partial_{r}\psi)^{2}]^{1/2}
r^{n-1}drd\tau,\label{eq5}
\end{equation}
where $r=\sqrt{x^{2}_{1}+\ldots+x^{2}_{n}}$. Although our main attention will be focused
on the cylindrical symmetry in BI electrodynamics ($n=2$ and $r=\sqrt{x^2+y^2}$),
we will also consider the spherical symmetry which corresponds to the case where
$n=3$ and $r=\sqrt{x^2+y^2+z^2}$. The generalization to the higher dimensions is
straightforward.

Equation~(\ref{eq4}) admits self-similar solutions of the form
\begin{equation}
\psi=ru(s),\quad s=\tau r^{-1}.
\label{eq6}
\end{equation}
Substituting Eq.~(\ref{eq6}) into Eq.~(\ref{eq4}) yields the
ordinary differential equation
\begin{equation}
(s^2-u^2-1)u''+(n-1)(u-su')[1-(u')^{2}+(u-su')^{2}]=0,
\label{eq7}
\end{equation}
where the prime denotes the derivative with respect to $s$.
We will seek an exact solution of Eq.~(\ref{eq7}) in
parametric form:
\begin{equation}
u=\xi^{1/2}\,\cosh\alpha,\quad s=\xi^{1/2}\,\sinh\alpha.
\label{eq8}
\end{equation}
Here, $\alpha=\int\Phi(\xi)d\xi+q$, where $q$ is an arbitrary integration constant.
Substituting expressions~(\ref{eq8}) into Eq.~(\ref{eq7}) and using the
formulas
\begin{equation}
u'_{s}=u'_{\xi}/s'_{\xi},\quad
u''_{ss}=(s'_{\xi}u''_{\xi\xi}-u'_{\xi}s''_{\xi\xi})/(s'_{\xi})^{3},
\label{eq9}
\end{equation}
we arrive at the Bernoulli equation
\begin{equation}
\frac{d\Phi}{d\xi}=-2\xi[(n-1)\xi-1]\Phi^{3}-
\frac{(4-n)\xi+3}{2\xi(\xi+1)}\Phi.
\label{eq10}
\end{equation}
Integration of Eq.~(\ref{eq10}) gives
\begin{equation}
\Phi=\pm {1\over 2}
\frac{(\xi+1)^{\gamma}}{\xi\sqrt{\chi_{n}(\xi)}},
\label{eq11}
\end{equation}
where $\gamma=(n-1)/2$, $\chi_{n}(\xi)=(\xi+1)^{n}-(p+n)\xi$ is a polynomial of order $n$,
and $p$ is
an integration constant. Restricting
ourselves to consideration only of the simplest cases $n=2$ and $n=3$, we write down
\begin{equation}
\chi_{2}(\xi)=\xi^{2}-p\xi+1
\label{eq12}
\end{equation}
and
\begin{equation}
\chi_{3}(\xi)=\xi^{3}+3\xi^{2}-p\xi+1.
\label{eq13}
\end{equation}
Thus, Eqs.~(\ref{eq6}), (\ref{eq8}), and~(\ref{eq11}) give an exact
solution of Eq.~(\ref{eq4}).
From these expressions, we have
\begin{eqnarray}
\partial_{\tau}\psi=\pm\frac{\sqrt{\chi_{n}(\xi)}\cosh\alpha+(\xi+1)^{\gamma}\sinh\alpha}
{\sqrt{\chi_{n}(\xi)}\sinh\alpha+(\xi+1)^{\gamma}\cosh\alpha},\nonumber\\
\partial_{r}\psi=\pm\frac{\sqrt{\xi}\,(\xi+1)^{\gamma}}
{\sqrt{\chi_{n}(\xi)}\sinh\alpha+(\xi+1)^{\gamma}\cosh\alpha}.
\label{eq14}
\end{eqnarray}
Now we should examine what physically meaningful solutions can be obtained by
appropriately choosing the arbitrary constants $p$ and $q$.

{\em Cylindrical symmetry.} Let us consider the following representation of the
quantity $\alpha$ in the case $n=2$:
\begin{equation}
\alpha=\pm {1\over 2} \int^{\xi}_{\xi_{2}}\frac{\sqrt{\xi+1}\,d\xi}
{\xi\sqrt{(\xi-\xi_1)(\xi-\xi_2)}},
\label{eq15}
\end{equation}
where $\xi_{1}=p/2-\sqrt{p^{2}/4-1}$, $\xi_{2}=p/2+\sqrt{p^{2}/4-1}$, $p>2$,
and $\xi>\xi_{2}$. Note that $\xi_1$ and $\xi_2$ are the real-valued roots of the polynomial $\chi_{2}(\xi)$
in Eq.~(\ref{eq12}).
Reduction of the elliptic integral in Eq.~(\ref{eq15}) to
the standard form gives
\begin{equation}
\alpha=\pm \frac{1}{\sqrt{\xi_{2}+1}} \left[(\xi_{1}-\xi_{2})\Pi(\beta,\nu,k)+
(\xi_{2}+1)F(\beta,k)\right],
\label{eq16}
\end{equation}
where $F$ and $\Pi$ are incomplete elliptic integrals of the first and
third kinds, respectively, $\beta=\sqrt{(\xi-\xi_{2})/(\xi-\xi_{1})}$,
$\nu=\xi_{1}/\xi_{2}$, and $k=\sqrt{(\xi_{1}+1)/(\xi_{2}+1)}$. We use
the following notation for the functions $\Pi$ and $F$:
\begin{equation}
\Pi(\beta,\nu,k)=\int^{\beta}_{0}\frac{d\zeta}{(1-\nu\zeta^2)
\sqrt{(1-\zeta^2)(1-k^2\zeta^{2})}},
\label{eq17}
\end{equation}
and $F(\beta,k)=\Pi(\beta,0,k)$.
The components of the axisymmetric electromagnetic field in BI electrodynamics
are given by Eqs.~(\ref{eq3}), (\ref{eq12}), (\ref{eq14}), and~(\ref{eq16}). With
the identity $ct/r=\xi^{1/2}\sinh\alpha$, these formulas determine $E$ and $B$
as functions of the radial coordinate and time via the parameter $\xi$ ($\xi_{2}<\xi<\infty$).

The energy conservation law $\partial_{t}W+\nabla\cdot{\bf \Sigma}=0$, with the
energy density
\begin{equation}
W=\frac{b^2}{4\pi}\left(\frac{1+b^{-2}B^{2}}{\sqrt{1-b^{-2}I}}-1\right)
\label{eq18}
\end{equation}
and the Poynting vector
\begin{equation}
{\bf \Sigma}=-{\hat {\bf e}_{r}}\frac{c}{4\pi}\frac{EB}{\sqrt{1-b^{-2}I}},
\label{eq19}
\end{equation}
can easily be derived directly from the field equations~(\ref{eq2}).
\begin{figure}[h]
\includegraphics{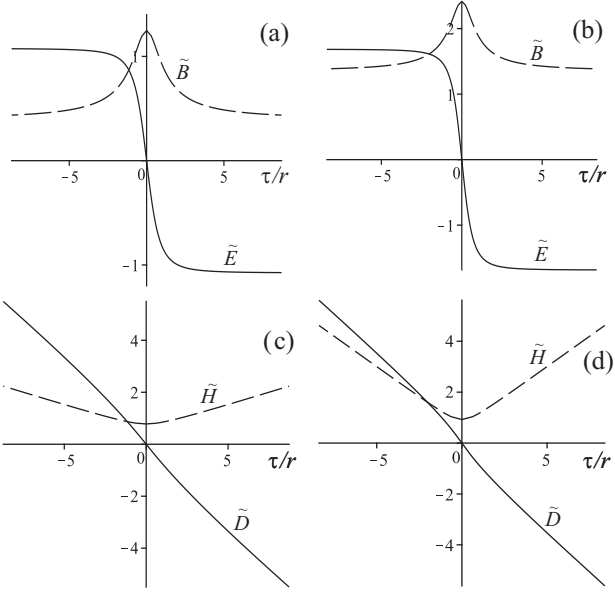}
\caption{Normalized field components $\tilde{E}=E/b$, $\tilde{B}=B/b$,
$\tilde{D}=D/b$, and $\tilde{H}=H/b$ as functions of $\tau/r$
for $p=2.2$ (a,c) and $p=6$ (b,d).}
\end{figure}
\begin{figure}[h]
\includegraphics{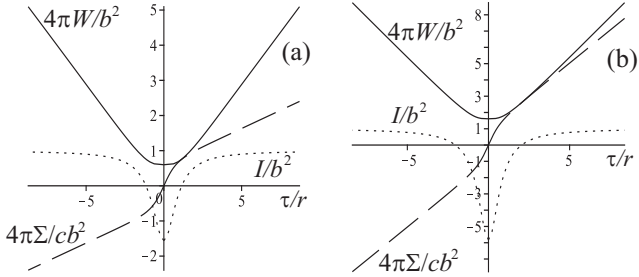}
\caption{Normalized energy density, the radial component of the Poynting vector, and the invariant $I$ ($4\pi W/b^2$, $4\pi \Sigma/cb^2$,
and $I/b^2$, respectively)
as functions of $\tau/r$ for $p=2.2$ (a) and $p=6$ (b). It is seen that $I/b^{2}\to 1$
for $\tau/r\to \pm\infty$.}
\end{figure}
\begin{figure}[h]
\includegraphics{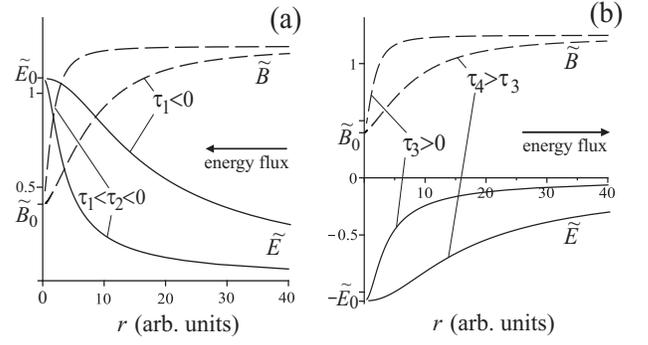}
\caption{Normalized electric and magnetic fields ($\tilde{E}=E/b$ and
$\tilde{B}=B/b$, respectively) as functions of $r$ for $p=2.2$ and various values of $\tau$:
$\tau_{1}<0$ and $\tau_{1}<\tau_{2}<0$ (a)
and $\tau_{3}>0$ and $\tau_{4}>\tau_{3}$ (b).}
\end{figure}
Figure~1 shows the results of calculations of $E$ and $B$ by
Eqs.~(\ref{eq3}) and~(\ref{eq14}) for $p=2.2$ and $p=6$. The quantities
$D$ and $H$ are also presented in this figure for the same values of $p$.
The plots of Fig.~1 can be considered as oscillograms of the field
quantities at a fixed point $r={\rm const}\neq 0$. The branch of
$\xi^{1/2}$ in Eq.~(\ref{eq8}) and the signs in Eq.~(\ref{eq14})
were chosen to provide the continuity of $E$ and $B$ at $\tau=0$ for
any $r\neq0$ and ensure that the radial component $\Sigma$ of the
Poynting vector is negative for $\tau<0$ and positive for $\tau>0$.
Figure~2 shows the
energy density $W$, the radial component $\Sigma$
of the Poynting vector, and the invariant $I$ as functions of $\tau/r$.
Figure~3 presents the snapshots of $E$ and $B$ as functions of $r$
for $p=2.2$ and various values of $\tau$. Since there is no characteristic spatial scale in
the problem considered, the radial coordinate is given in arbitrary units. It is
seen in Figs.~1 and 3 that the field quantities described by the obtained
solution are single-valued continuous functions at any space-time point,
except for the points on the symmetry axis. The solution describes the
propagation of a cylindrically symmetric  disturbance in the azimuthally
magnetized vacuum. For $\tau<0$, the cylindrical electromagnetic
wave converges to the axis and the wave profile becomes steeper [see Fig.~3(a)].
As a result, a shock wave forms at $r=0$ and $\tau=0$. At this time instant, the electric
field and the energy-flow direction reverse their signs, so that
for $\tau>0$ we observe propagation of a divergent cylindrical wave [see Fig.~3(b)].
Although the fields $E$ and $B$ are everywhere finite, it can be shown that
$D$, $H$, and $W$ behave as ${\rm const}\times r^{-1}$ for $r\to 0$ and $\tau\neq 0$.
Such a singularity is responsible for a linear increase in the dependences
$D(\tau/r)$, $H(\tau/r)$, $W(\tau/r)$, and $\Sigma(\tau/r)$ in Figs.~1(c), 1(d),
2(a), and 2(b). It is also seen in these
figures that (i) an increase in $W$ corresponds to the
positive values of the invariant $I$, (ii) the quantities $D$, $H$, and $W$
diverge not only at $\tau={\rm const}\neq 0$ and $r\to 0$, but also
at $r={\rm const}\neq 0$ and $\tau\to \infty$, and
(iii) the local group velocity $v_{g}=|\Sigma|/W$ does not exceed $c$.
The energy density is everywhere integrable.

The presence of the above singularity allows us to propose a physical interpretation
of the obtained exact solution as that due to a distributional source on the axis,
which is similar to static BIon solutions~\cite{Gib1,Gib2}. The divergent
cylindrical wave [see Fig.~3(b)] can be excited by ``switching-on'' of a
delta-function source on the axis at the time instant $\tau=0$. Since
charges and currents can not be specified independently of
the field in nonlinear
BI electrodynamics~\cite{Bor}, the existence of such a source is
inseparably related to the presence of a constant background azimuthal magnetic field
$B(r)\equiv B_{m}=b\sqrt{\xi_2}>b$ at $\tau=0$ and $r\neq 0$.
On the axis, the chosen source supports constant
fields $E(r=0,\tau>0)=-E_{0}$ and $B(r=0,\tau>0)=B_{0}$ such that
$I=E^{2}_{0}-B^{2}_{0}=b^{2}$ and ($D$, $H$)$\to\infty$. This limiting field state existing in
BI electrodynamics has received an interpretation in string theory
as a divergence in the rate of pair production of open strings~\cite{Bac,Gib2}.
An analogous effect follows from the QED model of vacuum polarization.
In QED, the electric field the strength of which is close to the Schwinger limit may cause
electron--positron pair creation from vacuum~\cite{Sch}.
The necessary condition for the Schwinger pair creation process is $I>0$~\cite{Sch,Mou}.
The interaction of
the created electrons and positrons with a
sufficiently strong field can lead to production
of multiple new particles and avalanche-like vacuum breakdown~\cite{Bul,Fed2}.
Thus, we can state that the obtained solution describes radial expansion of
the electron--positron plasma bunch. Due to an avalanche-like electromagnetic
cascade, we have infinite polarization and magnetization of vacuum. This process, which is
created on the axis, progressively propagates in the whole space. From the classical
viewpoint, such an expansion can be explained intuitively as a drift of
charged particles in the crossed fields $E_z$ and $B_\phi$.

{\em Spherical symmetry.} Since continuous tangential vector field on a sphere
cannot depend only on the radial coordinate, this case obviously has no direct
bearing on electrodynamics. However, the model of a spherically symmetric
scalar field with BI action~(\ref{eq5}) is studied intensely in
connection with string/M theory~\cite{Hop,Egg,Bro,Riz}.

In what follows, we consider the solution of Eq.~(\ref{eq4}) with $n=3$ in a way similar
to that used for the axial symmetry. For $n=3$, the constants $p$
and $q$ can be chosen so that the quantity $\alpha$ is given by
\begin{equation}
\alpha=\pm {1\over 2} \int^{\xi}_{\xi_{3}}\frac{(\xi+1)\,d\xi}
{\xi\sqrt{(\xi-\xi_1)(\xi-\xi_2)(\xi-\xi_3)}},
\label{eq20}
\end{equation}
where $\xi_1$, $\xi_2$, and $\xi_3$ are the real-valued roots of the polynomial $\chi_{3}(\xi)$
in Eq.~(\ref{eq13}) such that $\xi_3>\xi_2>\xi_1$.
Since complete analysis of the cubic equation is cumbersome, we consider
only a single value $p=5$ in Eq.~(\ref{eq13}). In this case, the above-mentioned roots
are equal to $\xi_{1}=-2-\sqrt{5}$, $\xi_{2}=-2+\sqrt{5}$, and $\xi_{3}=1$, so that
$\alpha$ can be represented as
\begin{equation}
\alpha=\pm \frac{\xi_{2}}{\sqrt{1-\xi_{1}}} \left[(\xi_{2}-1)\Pi(\beta,\xi_{2},k)+
(\xi_{2}+1)F(\beta,k)\right],
\label{eq21}
\end{equation}
where $\beta\!=\!\sqrt{(\xi-1)/(\xi-\xi_{2})}$
and $k\!=\!\sqrt{(\xi_{2}\!-\!\xi_{1})/(1-\xi_{1})}$.
The quantities $\partial_{\tau}\psi$ and $\partial_{r}\psi$ are given
by~Eqs.~(\ref{eq13}), (\ref{eq14}), and~(\ref{eq21}). The energy density and
flux can readily be found from Eqs.~(\ref{eq18}) and~(\ref{eq19})
by representing $E$ and $B$ in terms of the derivatives of $\psi$ in accordance
with Eq.~(\ref{eq3}).
\begin{figure}[h]
\includegraphics{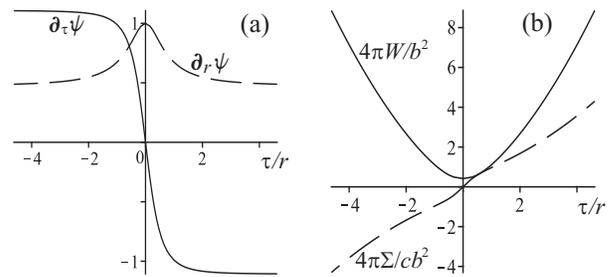}
\caption{Quantities $\partial_{\tau}\psi$ and $\partial_{r}\psi$ of
the spherically symmetric BI field (a) and the energy density and the radial component of the energy flux (b) as functions of $\tau/r$.}
\end{figure}
The results of calculations of $\partial_{\tau}\psi$, $\partial_{r}\psi$,
$W$, and $\Sigma$ are shown in Fig.~4. It is seen in the
figure that by analogy with the cylindrical case, the solution
describes the propagation of a spherically symmetric disturbance in
a constant background field. It can be shown that $W\sim{\rm const}\times r^{-2}$
for $r\to 0$ ($\tau\neq 0$) and, hence, the singularity at the origin
is integrable. The physical constraint $v_{g}=|\Sigma|/W\leq c$ is
also satisfied.

In conclusion, we emphasize that the obtained exact solutions exist
essentially due to BI nonlinearity and spatial symmetry. Maxwell equations ($b\to\infty$)
as well as BI equations in flat geometry [$n=1$ in Eq.~(\ref{eq4})] obviously
do not allow the existence of such solutions. Although the obtained solutions
do not have a finite total energy, their energy density is locally integrable.
Because of this, by suitably cutting off, the obtained solutions, as, e.g.,
plane wave solutions, may provide useful approximations
to solutions with finite total energy. Moreover, BI wave propagation in
a constant background field represents considerable interest
in string theory~\cite{Gib3}. Finally, we note that Eq.~(\ref{eq4}) also
admits self-similar solutions $\psi=\tau u(r/\tau)$. The parametrization
$u=\tanh\alpha$ and $r\tau^{-1}=(-\xi)^{-1/2}(\cosh\alpha)^{-1}$ with
$\alpha=\int\Phi(\xi)d\xi+q$ leads again to Eq.~(\ref{eq10}). Because
of discontinuity or ambiguity, we failed to give some of these solutions
any physical interpretation.
However, it can be assumed that among the whole set of
partial solutions, there may exist physically meaningful ones.
Since they are determined by three governing parameters ($n$ and two
integration constants) of the problem, an appropriate choice of these parameters needs
further studies.

This work was supported by the Government of the Russian Federation (Project No.~11.G34.31.0048),
the RFBR (Projects No.~12--02--00747-a and
No.~13--02--00234-a), and the Russian Ministry of Science and Education
(Contract No.~14.B37.21.1292).

\bibliography{Petrov_Kudrin_bibl}

\end{document}